\documentclass[aoas,MSNbibl,nameyear,rotating,seceqn,dvips]{arximspdf}
\usepackage{multirow,dcolumn}
\usepackage{graphicx}

\doi{10.1214/14-AOAS735} 
\volume{8}
\issue{2}
\pubyear{2014}
\firstpage{1232}
\lastpage{1255}

\makeatletter
\newcolumntype{d}[1]{D{.}{.}{#1}}
\newcommand{\rrvert}{\vert}
\newcommand{\llvert}{\vert}
\makeatother

\begin{document}
\begin{frontmatter}

\title{Gene-level pharmacogenetic analysis on survival outcomes using
gene-trait similarity regression}
\runtitle{Similarity regression for survival outcomes}

\begin{aug}
\author[a]{\fnms{Jung-Ying}~\snm{Tzeng}\thanksref{m1,m3,t1,t2}\ead[label=e1]{jytzeng@ncsu.edu}},
\author[b]{\fnms{Wenbin}~\snm{Lu}\thanksref{m1,t1,t3}\ead[label=e2]{lu@stat.ncsu.edu}}
\and
\author[c]{\fnms{Fang-Chi}~\snm{Hsu}\corref{}\thanksref{m2,t4}\ead[label=e3]{fhsu@wakehealth.edu}\ead[label=u1,url]{http://www.foo.com}}

\thankstext{t1}{Equal contribution.}
\thankstext{t2}{Supported by National Institutes of Health Grants R01
MH084022 and P01 CA142538.}
\thankstext{t3}{Supported by National Institutes of Health Grants R01
CA140632 and P01 CA142538.}
\thankstext{t4}{Supported by National Institutes of Health Grant U01 HG005160.}

\runauthor{J.-Y. Tzeng, W. Lu and F.-C. Hsu}

\affiliation{North Carolina State University\thanksmark{m1}, Wake
Forest University\thanksmark{m2}\break and National Cheng-Kung
University\thanksmark{m3}}

\address[a]{J.-Y. Tzeng\\
Department of Statistics\\
North Carolina State University\\
Raleigh, North Carolina 27695-8203\\
USA\\
and\\
Department of Statistics\\
National Cheng Kung University\\
Tainan City\\
Taiwan (R.O.C.)\\
\printead{e1}}

\address[b]{W. Lu\\
Department of Statistics\\
North Carolina State University\\
Raleigh, North Carolina 27695-8203\\
USA\\
\printead{e2}}

\address[c]{F.-C. Hsu\\
Department of Biostatistical Sciences\\
Wake Forest School of Medicine\\
Winston-Salem, North Carolina 27157\\
USA\\
\printead{e3}}
\end{aug}
%

\received{\smonth{3} \syear{2012}}
\revised{\smonth{1} \syear{2014}}


\begin{abstract}
Gene/pathway-based methods are drawing significant attention due to
their usefulness in detecting rare and common variants that affect
disease susceptibility. The biological mechanism of drug responses
indicates that a gene-based analysis has even greater potential
in pharmacogenetics.
Motivated by a study from the Vitamin Intervention for Stroke
Prevention (VISP) trial, we
develop a gene-trait similarity regression for survival analysis
to assess the effect of a gene or pathway on time-to-event outcomes.
The similarity
regression has a general framework that covers a range of survival
models, such as the proportional hazards model and the proportional
odds model. The inference procedure developed under the proportional
hazards model is robust against model misspecification. We
derive the equivalence between the similarity survival regression
and a random effects model, which further unifies the current
variance component-based methods. We demonstrate the effectiveness
of the proposed method through simulation studies. In addition,
we apply the method to the VISP trial data to
identify the genes that exhibit an association with the risk of a recurrent
stroke. The \emph{TCN2} gene was found to be associated with
the recurrent stroke risk in the low-dose arm. This gene may
impact recurrent stroke risk in response to cofactor therapy.
\end{abstract}

%

\end{frontmatter}

\section{Introduction}\label{sec1}

Genetic variations play a significant role in drug responses.
A gene that participates in a particular physiological mechanism might
influence the response to a specific therapeutic agent that targets the
mechanism. Identifying these influential genes may help to clarify if
an individual might benefit from or be harmed by the therapy.
Understanding the
genetic diversity of drug responses can help to identify
medications that maximize treatment effectiveness and minimize the
risk of adverse effects for individuals. Such an understanding will
also lead to
improved risk stratification, prevention and treatment strategies
for human diseases.
Pharmacogenetics studies show how an adverse reaction or positive
response to
pharmaceutical treatment is affected by an individual's genetic makeup
and has the potential to
deliver both public health and economic benefits rapidly. With the
recent advancements in
high-throughput technologies, it is becoming common for
pharmacogenetic researches to systematically investigate genetic
markers across the genome. Nevertheless, appropriate and
efficient analysis of the data remains a challenge.

Gene- or pathway-based analyses can assess pharmacogenetic effects more
effectively than single-marker based analyses
[\citet{GolTatSis03};
\citet{Gol05}]. First, there often exist obvious candidate genes
and pathways that metabolize the drug and carry variants
that are relevant to the drug responses. Responses to therapies usually
involve complex relationships between gene variants within the same
molecular pathway or functional gene set. When applied to
pharmacogenetic studies, gene- or
pathway-based methods might identify
multiple variants of subtle effects that are missed by single
marker-based methods.
Second, pharmacogenetic studies typically enroll only
a moderate number of patients, which limits the power of the association
detection. Gene-based
analyses have been shown to yield higher power than standard
single marker and haplotype analyses. This type of analysis can
particularly facilitate studies on rare-event drug responses, such as
adverse reactions, where it could take many years to collect a
sufficient number of samples to obtain adequate power for standard analyses.
In gene-based analyses, the association signals are aggregated across
variants, and the total number of tests is reduced; the amplification
of the association signals and the alleviation of the multiple
testing burdens result in improved power.

\begin{figure}

\includegraphics{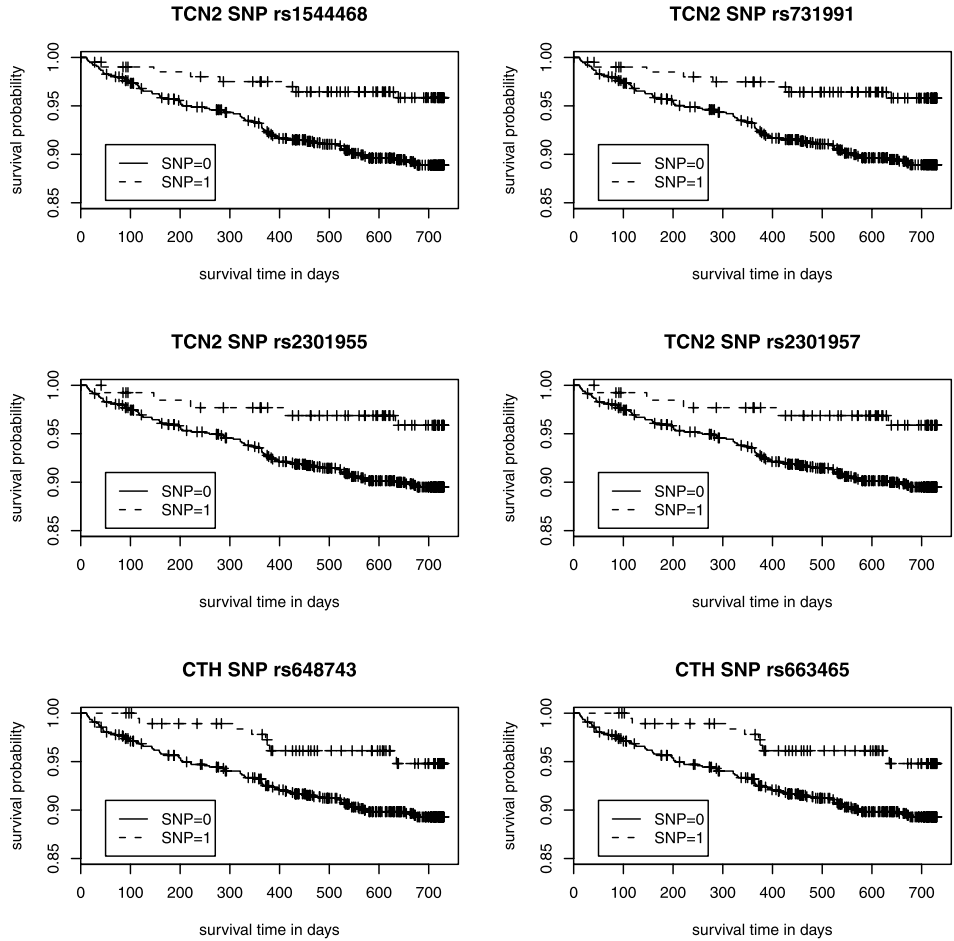}

\caption{The Kaplan--Meier survival curves for the top 6 SNPs
identified from the single SNP association analysis with risk of
recurrent stroke.}
\label{fig.top6}
\end{figure}

Our study was motivated by the need for a gene-based analysis of
the time-to-event data of the Vitamin Intervention for Stroke
Prevention (VISP) trial [Toole, Malinow and Chambless (\citeyear{TooMalCha04}); Hsu, Sides and Mychalecky
(\citeyear{HsuSidMyc11})].
Our
goal is to assess the association between the recurrent
stroke risk and the 9 candidate genes involved in the homocystein
(Hcy) metabolic pathway (see Data section for more details).
In our
preliminary analysis, we used the Cox proportional hazards (PH) model
[\citet{Cox72}] to perform single-SNP screening on 69 SNPs
across 9 genes from 969 individuals.
There were no SNPs past
the significance threshold after accounting for multiple testing.
However, the top 6 hits, thresholding at unadjusted $p$-values
$<$0.05, were concentrated in two genes. Specifically, 4 SNPs are
from \emph{TCN2} (i.e., rs1544468, rs731991, rs2301955 and
rs2301957 have Wald's test $p$-values of 0.0065, 0.0072, 0.0346 and 0.0346,
resp.) and 2 SNPs are from \emph{CTH} (i.e., rs648743 and
rs663465 each have a Wald's test $p$-value of 0.0115). The Kaplan--Meier
curves of these
6 SNPs are shown in Figure~\ref{fig.top6} and indicate the potential
for different risk
patterns among different variants at these loci.
The clustering
within the two genes suggests that it would be more efficient to
combine the individual signal strengths and model the joint effect of
multiple loci in a gene.

We perform the gene-based analysis using {a gene-trait
similarity regression inspired by Haseman--Elston
regression from linkage analysis [\citet{Elsetal00}; \citet{HasEls72}] and haplotype similarity tests for regional association
[\citet{Becetal05}; \citet{QiaTho01}; \citet{Tzeetal03}].}
{First, we quantify the genetic and trait similarities
for each pair of individuals. The genetic similarity is determined
using identity by state (IBS) methods. The trait similarity is obtained
from the covariance of the transformed survival time conditional on the
covariates.}
We then regress the trait similarity on
the genetic similarity and test the regression coefficient to detect
the genetic association.
There are several gene-based approaches
for censored time-to-event phenotypes in the literature, including
\citet{Goeetal05} and Lin and colleagues [\citet{CaiTonLin11}; \citet{Linetal11}]. In these approaches, the multimarker effects were
modeled under the Cox PH model using linear random effects [\citet{Goeetal05}] or a
nonparametric function induced by a kernel machine [\citet{CaiTonLin11}; \citet{Linetal11}].
The global
effect of a gene was detected by testing for the corresponding
genetic variance component. These approaches were found to be superior
in identifying pathways or genes that are associated
with survival.

{For many years, similarity-based methods have been successfully used to
evaluate gene-based associations} in quantitative and binary traits
[\citet{Becetal05}; \citet{LinSch09}; \citet{QiaTho01};
\citet{Tzeetal03}; \citet{WesSch06}]. Our work makes such
approaches available for survival phenotypes.
In addition, our similarity regression covers a variety of risk models,
including the commonly used PH model and the proportional odds (PO) model.
Furthermore, we show that the coefficient of the similarity
regression obtained for survival phenotypes can be reexpressed as a
variance component of a
certain working random effects model.
Such results facilitate the derivation of the
test statistic and unify the similarity model and previous
variance component methods
[\citet{Goeetal05}; \citet{CaiTonLin11}; \citet{Linetal11}].
{Specifically, under the Cox PH model, our test statistic
is equivalent to the test statistic defined by a kernel machine
approach [\citet{Linetal11}].}
We also show that the test statistic can be robust to model
misspecification. {Specifically, the proposed test gives the correct
type I error even if
the true risk model is misspecified.} However, the correct
specification of the true risk model generally leads to a test with
better power.
{Finally, we demonstrate the utility of the similarity
regression by identifying the important \emph{TCN2} gene in the VISP
study. The significance of \emph{TCN2} to stroke risk has been reported
by other association studies [\citet{Giuetal10}; \citet{Lowetal11}]
and has been supported by molecular biology evidence [\citet{Afmetal03}; \citet{Casetal05}]. Our findings further
suggest potential interactions between \emph{TCN2} and B12
supplementation. This new information furthers the possibility that
\emph{TCN2} could be utilized to predict recurrent strokes, identify
at-risk individuals and identify therapeutic targets for ischemic stroke.}

\section{Data}

The VISP study was a prospective, double-blind, randomized clinical trial
[Toole, Malinow and Chambless (\citeyear{TooMalCha04})]. The trial was
designed to study if high doses of
folic acid, vitamin B6 and vitamin B12\vadjust{\goodbreak} reduce the risk of a
recurrent stroke as compared to low doses of these vitamins.
The trial enrolled patients who were
35 or older, had a nondisabling cerebral infarction within 120 days of
randomization, and had Hcy levels in the top quartile
of the U.S. population. Subjects were randomly assigned to receive
daily doses of either a high-dose formulation (containing 25~mg
vitamin B{6}, 0.4~mg vitamin B{12} and 2.5~mg folic acid) or
a low-dose formulation
(containing 200~$\mu$g vitamin B{6}, 6~$\mu$g vitamin B{12} and
20~$\mu$g folic acid). Patient recruitment began in August 1997 and was
completed in December 2001. A total of 3680 participants were
enrolled at 56 clinical sites across the United States, Canada and
Scotland. The patients were followed for a maximum of two years,
and the average follow-up duration was 1.7 years. In the VISP genetic
study, 2206 participants provided informed consent and blood
samples. The SNP genotypes of 9 genes related to the
enzymes and cofactors in the Hcy metabolic pathway were collected: \emph
{BHMT1, BHMT2, CBS, CTH, MTHFR, MTR, MTRR, TCN1
}and \emph{TCN2} [Hsu, Sides and Mychalecky (\citeyear{HsuSidMyc11})]. In a
previous study, Hsu, Sides and Mychalecky (\citeyear{HsuSidMyc11}) conducted single-SNP analyses on targeted loci
(e.g., Hcy-associated variants) to examine the genetic association
with the recurrent stroke risk.
In the low-dose arm, the authors found that \emph{TCN2} SNP
rs731991 under a recessive mode was associated with the risk of
a recurrent stroke with an unadjusted log-rank test $p$-value of 0.009.
The associations for the
remaining SNPs within the 9 genes in the low-dose arm were not
studied. We extend this previous analysis to all 9 genes using a
gene-based approach. After quality control screening of
the data (e.g., removing loci with $>$99\% missing proportion or
Hardy--Weinberg disequilibrium under additive mode and removing
individuals with missing genotypes), the analysis included 969
individuals in the low-dose arm
with 69 recessively coded SNPs.

\section{Gene-trait similarity regression for survival traits}
\subsection{The model}
For individual $i$ $(i=1,2,\ldots,n)$, let $T_{i}$ denote the survival
time of
interest and $C_{i}$ denote the censoring time. We observe $\tilde
{T}_{i}=\min
( T_{i},C_{i} ) $ and the censoring indicator $\delta
_{i}=I(T_{i}\leq C_{i})$. In addition, let $X_{i}$ denote the $K\times
1$ vector
of covariates and \textsc{g}$_{mi}$ denote the allele count vector of
marker $m$
for person~$i$, where the length of \textsc{g}$_{mi}$, $\ell_{m}$ is
the number of distinct alleles at
marker $m$, $m=1,2,\ldots,M$. For example, for
a tri-allelic locus $m$, \textsc{g}$_{mi}=(1,0,1)^{T}$ if person $i$ has
genotype ``$A_{1}A_{3}$'' and $(0,2,0)^{T}$ if person $i$ has genotype
``$A_{2}A_{2}$.''

For each pair of individuals $i$ and $j$, we measure the genetic
similarity, $S_{ij}$, of the targeted gene and the trait similarity,
$Z_{ij}$. The genetic
similarity is quantified using the weighted IBS sum across the $M$
markers in
the gene, that is, $S_{ij}=\sum_{m=1}^{M}w_{m}S_{ij}^{m}$, where $S_{ij}^{m}=2$
if $\llvert\mbox{\textsc{g}}_{m,i}-\mbox{\textsc{g}}_{m,j}\rrvert$
is a zero vector, $S_{ij}^{m}=1$ if $\llvert\mbox{\textsc{g}}_{m,i}-
\mbox{\textsc{g}}_{m,j}\rrvert$ contains exactly two $1$'s (and
if $\ell_{m}>2$, the remaining entries are~$0$), and
$S_{ij}^{m}=0$ otherwise. The weights, $w_{m}$, are specified to
up-weight or down-weight a variant based on certain features. Examples
include weights
that are based on allele frequencies, the degree of evolutionary
conservation or the functionality of the variations
[\citet{WesSch06};
\citeauthor{Sch10N1} (\citeyear{Sch10N1,Sch10N2});
\citet{Prietal10}]. We
can use the minor
allele frequency of marker $m$, denoted as $q_{m}$,
to up-weight similarities that are contributed by rare variants.
Specifically, one can set a moderate weight, such as
$w_{m}=q_{m}^{-{3}/{4}}$ [Pongpanich, Neely and
Tzeng (\citeyear{PonNeeTze12})] or
$w_{m}=q_{m}^{-1}$ [\citet{Tzeetal11}], to promote similarity
attributed by rare
alleles, or use a more extreme weight, such as $w_{m}= (
1-q_{m} ) ^{24} $ [\citet{Wuetal11}], to target rare variants only.

The trait similarity, $Z_{ij}$, is quantified as follows. First, we
define $%
Y_{i}=H ( T_{i} ) $, where $H(\cdot)$ is an (unspecified) monotonic
increasing transformation function, such as the logarithm transformation
$Y_{i}=\log( T_{i} ) $. Assume that the conditional mean of $Y_{i}$
given the covariates and genes is $E ( Y_{i}\mid X_{i},
{g_{i}} )
=\theta+X_{i}^{T}\gamma+
{g_{i}}$, where $\theta$ is the intercept,
{$g_{i}$ is the multi-locus}
genetic effect of person $i$, and $\gamma$ is the $K$-dimensional
covariate effect. Further, define $\mu_{i}^{0}=\theta+X_{i}^{T}\gamma
$. The
trait similarity is defined as the product of the paired residuals adjusting
for the covariate effects, that is, $Z_{ij}=(Y_{i}-\mu
_{i}^{0})(Y_{j}-\mu
_{j}^{0})$. The expected value of the trait similarity is the
covariance between
the transformed survival times of subjects $i$ and $j$.

The gene-trait similarity regression has the form
%
\begin{equation}
E ( Z_{ij}\mid X_{i},X_{j} ) =b\times
S_{ij},\qquad i\neq j. \label{simreg}
\end{equation}
Just as in \citeauthor{Tzeetal09} (\citeyear{Tzeetal09,Tzeetal11}), the regression
has a zero intercept and does not have the covariate term
$X_{i}X_{j}$ because the baseline and covariate effects have been
adjusted when defining $Z_{ij}$. This argument will become more
obvious from the viewpoint of variance components in the following
subsection. Under model (\ref{simreg}), the overall association of
a gene can be evaluated by testing the null hypothesis: $b=0$.

\subsection{Score test for the gene-level effect}\label{scoretest}

We derive the score test statistic based on the equivalence between the
similarity regression and a mixed model. This equivalence is
demonstrated as follows.
Consider a working mixed model for the transformed survival time:
%
\begin{equation}
Y_{i}=H(T_{i})=X_{i}^{T}
\gamma+{g_{i}}+\theta+\varepsilon_{i}, \label{VCmodel}
\end{equation}
where
$
{(g_{1},\ldots,g_{n})}^{T}\sim N$ $ ( 0,\tau S ) $ with $S= \{
S_{ij} \} _{i,j=1}^{n}$, that is, the covariance between
{$g_{i}$ and $g_{j}$}
depends on the genetic similarity between subjects $i$ and $j$, and $%
\varepsilon_{i}^{\ast}\equiv\theta+\varepsilon_{i}$,
$i=1,\ldots,n$ are independently and identically distributed with a
known distribution that is independent of $X_{i}$ and ${g_{i}}$. Given
$X_{i}$ and {$g_{i}$}, model~(\ref
{VCmodel}) specifies a
general class of linear transformation models [\citet{CheWeiYin95}], which contains many popular survival models as special
cases. For example, when $\varepsilon_{i}^{\ast}$ follows the
standard extreme value distribution, the linear transformation model
becomes the PH model [\citet{Cox72}]. When $\varepsilon_{i}^{\ast}$ follows the standard logistic
distribution, the linear transformation model becomes the PO model
[Bennett (\citeyear{Ben})].

Under (\ref{VCmodel}), the conditional expectation of the trait similarity
between individuals $i$ and $j$ ($i\neq j$) is
\begin{eqnarray*}
\label{covY} E(Z_{ij}\mid X_{i},X_{j}) &=&
\operatorname{cov} ( Y_{i},Y_{j}\mid X_{i},X_{j}
) =\operatorname{cov}\bigl(g_{i}+\varepsilon_{i}^{\ast
},g_{j}+
\varepsilon_{i}^{\ast}\bigr)
\\
&=&\operatorname{cov}(g_{i},g_{j})
\\
&=&\tau\times S_{ij}.
\end{eqnarray*}
Therefore, we have $b=\tau$, that is, the regression coefficient in the
similarity regression (\ref{simreg}) is the genetic variance component in
the mixed model (\ref{VCmodel}). This motivates us to develop a score test
for the variance component in the working model. As shown in the
\hyperref[app]{Appendix},
the score test statistics for $\tau=0$ can be written as
\[
Q_{n}=\frac{1}{n}(\hat{r}_{1},\ldots,
\hat{r}_{n})S(\hat{r}_{1},\ldots,\hat{r}%
_{n})^{T},
\]
where
\begin{eqnarray*}
\hat{r}_{i}&=&\int_{0}^{\infty}\hat{
\omega}_{i}(t)\,dM_{i}(t;\hat{\gamma},\hat{H})\\
& =&
\delta_i\frac{\dot{\lambda} \{\hat{H}(\tilde{T}_i)-\hat
{\gamma}^TX_{i}\}}{\lambda\{\hat{H}(\tilde{T}_i)-\hat{\gamma}^TX_{i}\}
} - \lambda\bigl\{\hat{H}(
\tilde{T}_i)-\hat{\gamma}^TX_{i}\bigr\},
\\
M_{i}(t;\gamma,H)&=&\delta_{i}I(\tilde{T}_{i}\leq
t)-\int_{0}^{t}I(\tilde{T%
}_{i}\geq s)\,d\Lambda\bigl\{H(s)-\gamma^TX_{i}
\bigr\},
\end{eqnarray*}
$\hat{%
\omega}_{i}(t)=\dot{\lambda}\{\hat{H}(t)-\hat{\gamma}^TX_{i}\}/\lambda
\{\hat{H}(t)-\hat{\gamma}^TX_{i}\}$, and $S$ is as defined after
equation~(\ref{VCmodel}). Here, $\lambda
(\cdot)$ and $\Lambda(\cdot)$ are the hazard and cumulative hazard
functions of $\varepsilon_{i}^{\ast}$, respectively, $\dot{\lambda
}(\cdot)$ is the first derivative of $\lambda(\cdot)$, and $\hat
{\gamma}$ and $\hat{H}%
(\cdot)$ are the estimates of $\gamma$ and $H(\cdot)$, respectively, in
model (\ref{VCmodel}) under the null hypothesis: $\tau=0$.
{For example, if the PH model is
imposed, that is, $\lambda(u) = \dot{\lambda}(u) = e^u$, the estimators
$\hat{%
\gamma}$ and $\hat{\Gamma}(\cdot)\equiv e^{\hat{H}(\cdot)}$ can be taken
as the maximum partial likelihood estimator and Breslow's estimator,
respectively.} Under this case, $\hat{\omega}_{i}(t)\equiv1$ and $\hat
{r}%
_{i}=\delta_{i}-\hat{\Gamma}(\tilde{T}_{i})\exp(-\hat{\gamma}^TX_{i})$,
that is, the martingale residual for the null model. {If
the PO model is used, that is, $\lambda(u) = e^u/(1+e^u)$ and $\dot
{\lambda}(u) = e^u/(1+e^u)^2$, we have $\hat{\omega}_{i}(t) = 1/[1+\exp
\{\hat{H}(t)-\hat{\gamma}^TX_i\}]$.
In general, $\gamma$ and $%
H(\cdot)$ can be estimated using the martingale-based estimating
equations [Chen, Jing and Ying (\citeyear{CheJinYin02})]} or the nonparametric maximum
likelihood estimation method [\citet{ZenLin06}] for the
semiparametric linear
transformation model.
In the \hyperref[app]{Appendix}, we show that under the null
hypothesis the test statistic, $Q_{n}$, asymptotically follows a
weighted $%
\chi^{2}$ distribution where the weights can be estimated
consistently. The $p$-values can then be calculated numerically using
a resampling method or moment-matching approximations [\citet{Pea59}; Duchesne and Lafaye
De~Micheaux (\citeyear{DucLaf10})].

{We note that the proposed score test can be
robust to the misspecification of the true survival model. As an
illustration, consider the test derived under the PH model. Based on
the results for the robust inference of the
PH model [\citet{LinWei89}], when the PH model is misspecified,
the maximum
partial likelihood estimator, $\hat{\gamma}$, and Breslow's estimator,
$\hat{%
\Gamma}(\cdot)$, do not converge to the true parameters but converge to
some deterministic values $\gamma^{\ast}$ and $\Gamma^{\ast}(\cdot
)$ under certain regularity conditions.}
The corresponding $r_{i}^{\ast}=\delta_{i}-\Gamma^{\ast}(\tilde{T}%
_{i})\exp(-X_{i}^{T}\gamma^{\ast})$ is not a martingale residual but has
a mean of 0 under the null hypothesis. Therefore, it can be shown that $Q_{n}$
converges in distribution to a weighted $\chi^{2}$ distribution under
the null hypothesis
even with model misspecification. {As shown in our simulation
studies, the proposed score test derived under the PH model still gives
correct type-I error when
the true model is the PO model. However, the power of the test depends
on the assumed working model. In general, the test derived under the
true risk model may have better power.}

\section{Results of the analysis of the VISP trial data}

We now return to the VISP trial to evaluate the
association between the recurrent stroke risk and the 9 candidate
genes studied in Hsu, Sides and Mychalecky (\citeyear{HsuSidMyc11}). In our analysis, we conducted a
gene-based screening on the 9 genes using the low-dose samples.
After removing loci with $>$1\% of missingness and
subjects with missing genotypes, there were 969 individuals
with 69 polymorphic SNPs under recessive coding.
Of the 969
individuals, 86 experienced a recurrent stroke (i.e., 91.1\%
censoring). We used the proposed
similarity regression (referred to as SimReg) with inverse allele
frequency weights
{$w_{m}=q_{m}^{-{3}/{4}}$, that is, the weight
recommended in Pongpanich, Neely and
Tzeng (\citeyear{PonNeeTze12}) when analyzing a mixture of common and rare variants.} We
calculated the $p$-values of the SimReg statistics
using the resampling method. Specifically, we computed the nonzero
eigenvalues, $\hat{\xi}_{1},\ldots,\hat{\xi}_{d}$, of
$\hat{\Sigma} $ as defined in the \hyperref[app]{Appendix}. We
generated $10^4$
sets of $ ( \chi_{1,1}^{2},\ldots,\chi_{1,d}^{2} ) $.
Each set consisted of $d$ independently and identically distributed
$\chi_{1}^{2}$ random variables. For each set, we calculated the
value $\sum_{k=1}^{d}\hat{\xi}_{k}\chi_{1,k}^{2}$, and the $10^4$
values formed an empirical null distribution of the SimReg
statistics. The SimReg $p$-value was the proportion of the generated
null statistics that were greater than the observed statistic.
{We performed SimReg analyses under the PH model
(referred to as SimReg-PH) and the PO model (referred to as SimReg-PO).}
The
performances of the SimReg methods were benchmarked against {three}
approaches: (a)
the single SNP minimum $p$-value method using the Cox PH model (referred
to as minP), (b) the {multi-SNP} method using the global
test for
survival under the PH model [\citet{Goeetal05}] as implemented in
the R-package
\texttt{globaltest} (referred to as Global),
{and (c) the multi-SNP method using the kernel machine [\citet{Linetal11}] as implemented in the R-package \texttt{KMTest.surv} (referred
to as KM) with $10^4$ resamplings. Although the SimReg-PH test
statistic is identical to the KM test statistic, the results may be
slightly different due to the different resampling methods adopted to
obtain the $p$-values. Specifically, in the KM method, the score
statistic was perturbed by multiplying i.i.d. normal random variables
to achieve the same limiting distribution of the test statistic. In
SimReg-PH, the weights in the limiting weighted $\chi^2$ distribution,
that is, ${\xi}_{1},\ldots,{\xi}_{d}$, were consistently estimated and
then the samples were directly generated from the estimated weighted
$\chi^2$ distribution based on a large number of i.i.d. $\chi_1^2$
random variables. The different resampling approaches also lead to
different computational burden. For example, using a 3.6 GHz Xeon
Processor with 60~GB RAM with $10^4$ resamplings, the system run-time of
SimReg-PH was $<\!1/6$ of KM in the VISP analysis, and the time
difference became greater with a larger number of resamplings.}
For the minP
method, we fitted the standard PH model to each SNP in a gene, took the
smallest $p$-value and
calculated the adjusted $p$-value of a gene to correct for the
multiple SNPs using $1-(1-$minimum raw
$p$-value)$^{K_{\mathrm{eff}}}$. The effective number of
independent tests, $K_{\mathrm{eff}}$, was estimated using the method of
\citet{MosSch08} and accounts for the correlations in recessive coding of
loci. As studied by Hsu, Sides and Mychalecky (\citeyear{HsuSidMyc11}), all analyses were considered under
the recessive mode and were adjusted for age, sex and race.

\begin{table}
\tabcolsep=0pt
\caption{Results of the VISP genetic study}\label{tab1}
\begin{tabular*}{\textwidth}{@{\extracolsep{\fill}}lccccccccc@{}}
\hline
& \textbf{\textit{BHMT1}} & \textbf{\textit{BHMT2}} &
\textbf{\textit{CBS}} & \textbf{\textit{CTH}} & \textbf{\textit{MTHFR}}
& \textbf{\textit{MTR}} & \textbf{\textit{MTRR}} & \textbf{\textit{TCN1}} &
\textbf{\textit{TCN2}} \\
\hline
Numbers of SNPs & \multicolumn{1}{c}{5} & \multicolumn{1}{c}{3} &
\multicolumn{1}{c}{6} & \multicolumn{1}{c}{10} & \multicolumn{1}{c}{7} &
\multicolumn{1}{c}{20} & \multicolumn{1}{c}{5} & \multicolumn{1}{c}{3} &
\multicolumn{1}{c}{15} \\
minP & \multicolumn{1}{r}{0.3399} & \multicolumn{1}{r}{0.5968} &
\multicolumn{1}{r}{0.3354} & \multicolumn{1}{r}{0.0918} &
\multicolumn{1}{r}{0.9105} & \multicolumn{1}{r}{0.8933} & \multicolumn{1}{r}{0.6183} &
\multicolumn{1}{r}{0.9764} & \multicolumn{1}{r}{0.0704} \\
($K_{\mathrm{eff}}$) & \multicolumn{1}{c}{(3.99)} & \multicolumn{1}{c}{(2.83)}
&
\multicolumn{1}{c}{(4.41)} & \multicolumn{1}{c}{(8.35)} & \multicolumn{1}{c}{
(4.64)} & \multicolumn{1}{c}{(10.64)} & \multicolumn{1}{c}{(4.78)} &
\multicolumn{1}{c}{(2.77)} & \multicolumn{1}{c}{(11.28)} \\
Global & \multicolumn{1}{r}{0.6457} & \multicolumn{1}{r}{0.7391} &
\multicolumn{1}{r}{0.2669} & \multicolumn{1}{r}{0.0518} & \multicolumn{1}{r}{
0.7819} & \multicolumn{1}{r}{0.9154} & \multicolumn{1}{r}{0.7363} &
\multicolumn{1}{r}{0.9689} & \multicolumn{1}{r}{0.0457} \\
{KM} & \multicolumn{1}{r}{0.4863} & \multicolumn
{1}{r}{0.6142} &
\multicolumn{1}{r}{0.2386} & \multicolumn{1}{r}{0.0078} & \multicolumn{1}{r}{
0.7094} & \multicolumn{1}{r}{0.8289} &
\multicolumn{1}{r}{0.7289} &
\multicolumn{1}{r}{1.0000} & \multicolumn{1}{r}{0.0075} \\
SimReg-PH & \multicolumn{1}{r}{0.5794} & \multicolumn{1}{r}{0.6402} &
\multicolumn{1}{r}{0.1889} & \multicolumn{1}{r}{0.0073} & \multicolumn{1}{r}{
0.6833} & \multicolumn{1}{r}{0.8835} & \multicolumn{1}{r}{0.5988} &
\multicolumn{1}{r}{0.9845} & \multicolumn{1}{r}{0.0040} \\
SimReg-PO & \multicolumn{1}{r}{0.6136} & \multicolumn{1}{r}{0.6942} &
\multicolumn{1}{r}{0.2877} & \multicolumn{1}{r}{0.0075} & \multicolumn{1}{r}{
0.7807} & \multicolumn{1}{r}{0.9011} & \multicolumn{1}{r}{0.6422} &
\multicolumn{1}{r}{0.9922} & \multicolumn{1}{r}{0.0052} \\ \hline
\end{tabular*}
\tabnotetext[]{}{The adjusted $p$-values for the gene were obtained using
$1- ( 1-\mbox{minimum raw $p$-value}) ^{K_{\mathrm{eff}}}$. Significance
is concluded by
comparing the $p$-values with the Bonferroni threshold $0.05/9=0.0056$,
which
accounts for the 9 gene analysis.}
\end{table}

The $p$-values for each of the methods are shown in Table~\ref{tab1}, and the
$p$-values are compared with the Bonferroni threshold adjusted for
the 9 gene analyses, that is, $0.05/9=0.0056$.
SimReg-PH detected
a significant association between the recurrent stroke risk and
\emph{TCN2} (i.e., $p$-value${} = 0.0040$), which strengthens the
observation of differential survival between different variants from
the single SNP analysis. Gene \emph{CTH} had the second smallest
$p$-value (0.0073), which did not pass the Bonferroni threshold but
was near the cutoff. These results also agree with the findings in the
single SNP analysis.
The results of SimReg-PO are similar to those of SimReg-PH except that
the $p$-values are slightly larger, that is, $p$-value${} = 0.0052$ for \emph
{TCN2} and 0.0075
for \emph{CTH}.
{On the other hand, the KM,} Global and minP methods did
not yield any significant findings. However, the smallest $p$-values
of the 9 genes were obtained for \emph{TCN2} (i.e., $p$-values of
\emph{TCN2} are {0.0075 for KM}, 0.0457 for Global and
0.0704 for minP). As expected, {the results of KM were
very similar to those of SimReg-PH, except that the $p$-value of \emph
{TCN2} was slightly above the 0.0056 threshold.} The smallest $p$-values
for the minP method were from
the \emph{TCN2} SNP rs731991 with a raw Wald's test $p$-value of
0.0065. However, neither the raw minimum $p$-value (0.0065) nor the
adjusted $p$-value (0.0704) survived the significance threshold corrected for
multiple testing (0.0056). {All} methods
also had \emph{CTH} as the gene with the second smallest $p$-value
(i.e., $p$-values {0.0078 for KM,} 0.0518 for Global, and
0.00918 for minP).

\begin{figure}

\includegraphics{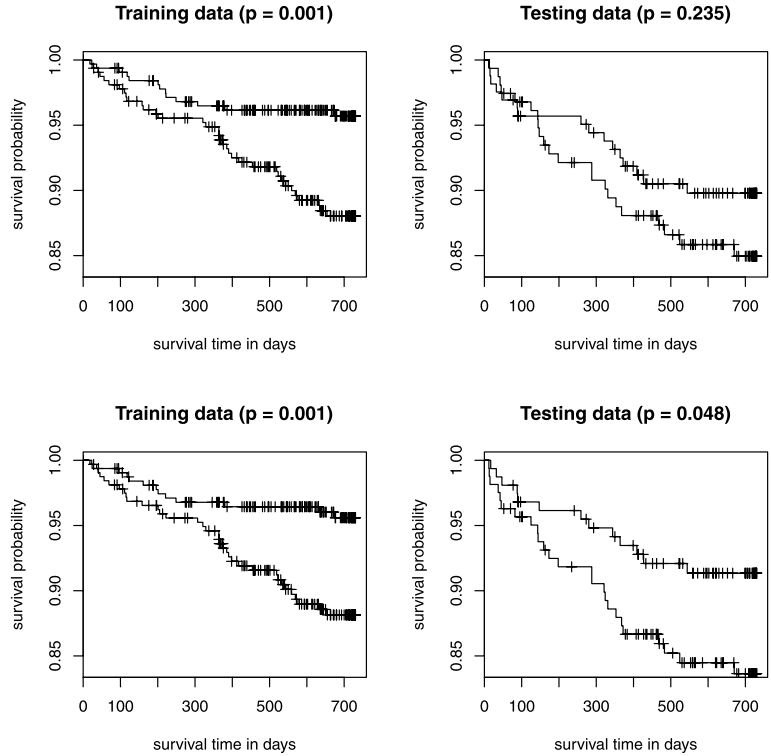}

\caption{The Kaplan--Meier survival curves for the two risk groups of patients.}
\label{fig.tcn2pred}
\end{figure}

Next, we assessed the prediction performance of Cox PH models built
with and without \emph{TCN2} using the procedure described in \citet{LiLua05}. Specifically, we randomly divided the samples into a
training set ($n= 646$) and a testing set ($n= 323$). Based on the
training set, we fitted the Cox PH regression with two models: Model 1
included only the baseline covariates (age, sex and race), that is, no
genetic information, and model 2 included the baseline covariates plus
the top 7 principal components (PCs) of \emph{TCN2} SNPs that explained
95\% of the variations. The PCs were used instead of the 15 original
genotypes because of the high linkage disequilibrium among them. Based
on the fits of the PH model from the training set, we obtained the risk
scores of every subject under each model and computed the medians of
the risk scores. We also computed the risk scores for the testing set
using the estimated coefficients from the training set. Next, we
divided the subjects into 2 risk groups: high-risk and low-risk.
Individuals with a risk score higher/lower than the median risk scores
obtained from the training data comprised the high-risk/low-risk group.
Finally, we plotted the Kaplan--Meier curves for the 2 risk groups in
the training data and the two risk groups in the testing data
separately and obtained the $p$-values of the corresponding log-rank
tests. The results are given in Figure~\ref{fig.tcn2pred}. As expected,
the $p$-values for both models under the training set were very
significant. However, only model 2 is significant ($p$-value is 0.048)
under the testing set. This result implies that \emph{TCN2} gives a
more accurate prediction of the risk for recurrent stroke.

Vitamin supplements have been identified as a potential treatment for
vascular diseases. The beneficial effects of vitamin supplements on
stroke recurrence are not yet fully understood. In VISP, vitamin
supplements did not show an effect on the recurrent stroke risk during
the 2 years of follow-up. However, we found that genetic variants, such
as SNPs in \emph{TCN2}, were associated with the recurrent stroke risk
in the low-dose arm. This finding is consistent with the literature.
\emph{TCN2} was previously found to be associated with ischemic stroke
risk [\citet{Lowetal11}; \citet{Giuetal10}] and premature ischemic
stroke risk [\citet{Giuetal10}]. It has been reported that \emph
{TCN2} interferes with the intracellular availability of vitamin B12
[\citet{Casetal05}]. The gene is associated with plasma
homocysteine levels and affects the proportion of vitamin B12 bound to
transcobalamin [\citet{Afmetal03}].
It is suspected that SNPs on the genes coding for enzymes involved in
the methionine metabolism have been suspected to be associated with
hyperhomocysteinaemia, which can result in occurrence of stroke [\citet{Giuetal10}].
Our significant findings in the low-dose arm suggest that there may be
an interaction between \emph{TCN2} and B12 supplementation, a finding
that warrants further studies. The findings lead to a hypothesis that
there may be one specific combination of genotypes of \emph{TCN2} that
is more efficient at transporting B12 and thus impacts the
effectiveness of cofactor therapy on recurrent stroke risk. A
functional study is being planned to localize possibly independent
regions of association and determine their function.

{Besides \emph{TCN2},} \emph{CTH} is marginally
associated with recurrent stroke risk in the low-dose arm. It encodes
cystathionine gamma-lyase, which is an enzyme that converts
cystathionine to cysteine in the trans-sulfuration pathway [\citet{Wanetal04}]. It may be a determinant of plasma Hcy concentrations, which
may increase the risk of recurrent stroke because of arterial disease.
{It is worth further study as well.}

\section{Simulation studies}

We performed simulations to assess the validity and effectiveness of the
proposed SimReg methods based on the 15 SNPs in \emph{TCN2} of the 969 VISP
low-arm samples. The rarest minor allele frequency (MAF) is
approximately 3\%.
We generated genotypes of $n$
individuals by randomly sampling with replacement from the 15-SNP genotypes
of the 969 samples.
For individual $i$, we generated the covariate, $X_{i}$, from $N (
0,1 ) $.
{We generated the survival time, $T_{i}$, based on the genetic and
covariate information under two models: the PH model and the PO model.
{Specifically, for the PH model,
$\log(T_{i})=-(X_{i}+\sum_{\ell} \gamma_{\ell}\mathcal{G}_{\ell
i})+\varepsilon_{i}^{\ast}$, where $\varepsilon_{i}^{\ast}$
follows the standard extreme value distribution;
for the PO model,
$\log\{\exp(T_{i})-1\}=-(X_{i}+\sum_{\ell}\gamma_{\ell}\mathcal
{G}_{\ell i})+\varepsilon_{i}^{\ast}$, where $\varepsilon_{i}^{\ast}$
follows the
standard logistic distribution.}
The value of $\mathcal{G}_{\ell i}$ is determined by the genotypes at
the causal locus $\ell$ and the mode of
inheritance. For example, if \textit{A} is the causal allele at locus
$\ell
$, then $\mathcal{G}_{\ell i}= 2, 1$ and 0 for genotypes \textit{AA, Aa}
and \textit{aa}, respectively, under an additive
mode. Under a dominant mode, $\mathcal{G}_{\cdot i}= 1, 1$ and 0,
respectively. Under a recessive mode, $\mathcal{G}_{\cdot i}= 1, 0$ and
0, respectively.
For type I error analysis, no SNPs were set to be causal, that is,
$\gamma_{\ell}$ was set to be 0 for all $\ell$.
For power analysis,
we selected 3 SNPs with different MAFs and LD patterns as causal loci
from the 15 SNPs in \emph{TCN2} and referred to them as SNP R, SNP U
and SNP~C. The MAFs are 0.036 (rare) for SNP R, 0.132 (uncommon) for
SNP U and 0.419 (common) for SNP~C. The average $R^2$'s between a causal
locus and the remaining loci are 0.002 (low) for SNP R, 0.003 (low) for
SNP U and 0.216 (high) for SNP~C. The specific values of $\gamma_{\ell
}$'s are given in Table~\ref{tab2} for each scenario under different inheritance
modes and censoring rates. The values were set to consider 1, 2 and 3
causal loci in the gene and to consider causal loci that have either
the same or different effect sizes (e.g., rarer variants with larger
effect sizes).}
All of the power scenarios assumed linear additive effects of the
causal loci, which favors the linear random effects model (e.g.,
Global) and can be used to examine the utility of using the nonlinear
IBS function to capture the multi-marker effects.

%
\begin{table}
\tabcolsep=0pt
\caption{Effect sizes for power analyses}\label{tab2}
\begin{tabular*}{\textwidth}{@{\extracolsep{\fill}}lcccccccc@{}}
\hline
& & \multicolumn{3}{c}{\textbf{Additive and dominant}} & &
\multicolumn{3}{c}{\textbf{Recessive}} \\ [-6pt]
& & \multicolumn{3}{c}{\hrulefill} & &
\multicolumn{3}{c@{}}{\hrulefill} \\
& & \textbf{15\%}$\bolds{^{\ast}}$ & \textbf{40\%} & \textbf{90\%} & &
\textbf{15\%} & \textbf{40\%} & \textbf{90\%} \\
& & $\bolds{n=500}$ & $\bolds{n=500}$ & $\bolds{n=1000}$ & &
$\bolds{n=1000}$ & $\bolds{n=1000}$ & $\bolds{n=1000}$ \\
\textbf{Scenario} & \textbf{Causal SNPs (MAF)} &
\multicolumn{3}{c}{\textbf{Effect size (}$\bolds{\gamma
_R,\gamma_U,\gamma_C}$\textbf{)}} & &
\multicolumn{3}{c}{\textbf{Effect size (}$\bolds{\gamma
_R,\gamma_U,\gamma_C}$\textbf{)}} \\
\hline
\phantom{0}1 & R (0.036) & \multicolumn{3}{c}{(1.5,\ 0.0,\ 0.0)} & & \multicolumn
{3}{c}{(4.0,\ 0.0,\ 0.0)} \\
\phantom{0}2 & U (0.132) & \multicolumn{3}{c}{(0.0,\ 1.0,\ 0.0)} & & \multicolumn
{3}{c}{(0.0,\ 3.0,\ 0.0)} \\
\phantom{0}3 & C (0.419) & \multicolumn{3}{c}{(0.0,\ 0.0,\ 0.3)} & & \multicolumn
{3}{c}{(0.0,\ 0.0,\ 0.3)} \\
\phantom{0}4 & R, U & \multicolumn{3}{c}{(0.6,\ 0.6,\ 0.0)} & & \multicolumn
{3}{c}{(4.0,\ 4.0,\ 0.0)} \\
\phantom{0}5 & R, U & \multicolumn{3}{c}{(0.6,\ 0.4,\ 0.0)} & & \multicolumn
{3}{c}{(2.5,\ 2.0,\ 0.0)} \\
\phantom{0}6 & R, C & \multicolumn{3}{c}{(0.3,\ 0.0,\ 0.3)} & & \multicolumn
{3}{c}{(0.3,\ 0.0,\ 0.3)}
\\
\phantom{0}7 & R, C & \multicolumn{3}{c}{(0.6,\ 0.0,\ 0.2)} & & \multicolumn
{3}{c}{(2.5,\ 0.0,\ 0.2)}
\\
\phantom{0}8 & U, C & \multicolumn{3}{c}{(0.0,\ 0.3,\ 0.3)} & & \multicolumn
{3}{c}{(0.0,\ 0.3,\ 0.3)}
\\
\phantom{0}9 & U, C & \multicolumn{3}{c}{(0.0,\ 0.4,\ 0.2)} & & \multicolumn
{3}{c}{(0.0,\ 2.0,\ 0.2)} \\
10 & R, U, C & \multicolumn{3}{c}{(0.3,\ 0.3,\ 0.3)} & & \multicolumn{3}{c}{
(0.3,\ 0.3,\ 0.3)} \\
11 & R, U, C & \multicolumn{3}{c}{(0.6,\ 0.4,\ 0.2)} & & \multicolumn{3}{c}{
(2.5,\ 2.0,\ 0.2)} \\
\hline
\end{tabular*}
\tabnotetext[]{}{$^{\ast}$censoring proportion.}
\end{table}

\begin{sidewaystable}
\tablewidth=\textwidth
\tabcolsep=0pt
%
\caption{Type I error rates for survival time generated from
the PH
model}\label{tab3}
\begin{tabular*}{\textwidth}{@{\extracolsep{\fill}}ld{1.2}d{1.2}d{1.2}c
d{1.2}d{1.2}d{1.2}cd{1.2}d{1.2}d{1.2}@{}}
\hline
& \multicolumn{3}{c}{\textbf{Additive}} & &
\multicolumn{3}{c}{\textbf{Dominant}} &&
\multicolumn{3}{c}{\textbf{Recessive}}\\[-6pt]
& \multicolumn{3}{c}{\hrulefill} &&
\multicolumn{3}{c}{\hrulefill} &&
\multicolumn{3}{c@{}}{\hrulefill}\\
\textbf{Analyzed under} & \multicolumn{1}{c}{\textbf{15\%}} & \multicolumn{1}{c}{\textbf{40\%}} &
\multicolumn{1}{c}{\textbf{90\%}} & & \multicolumn{1}{c}{\textbf{15\%}} &
\multicolumn{1}{c}{\textbf{40\%}} & \multicolumn{1}{c}{\textbf{90\%}} & &
\multicolumn{1}{c}{\textbf{15\%}} & \multicolumn{1}{c}{\textbf{40\%}} &
\multicolumn{1}{c@{}}{\textbf{90\%}}\\
\textbf{PH model} & \multicolumn{1}{c}{$\bolds{(n=500)}$} &
\multicolumn{1}{c}{$\bolds{(n=500)}$} &
\multicolumn{1}{c}{$\bolds{(n=1000)}$} & &
\multicolumn{1}{c}{$\bolds{(n=500)}$} &
\multicolumn{1}{c}{$\bolds{(n=500)}$} &
\multicolumn{1}{c}{$\bolds{(n=1000)}$} & &
\multicolumn{1}{c}{$\bolds{(n=1000)}$} &
\multicolumn{1}{c}{$\bolds{(n=1000)}$} &
\multicolumn{1}{c@{}}{$\bolds{(n=1000)}$} \\
\hline
\multicolumn{12}{c}{$\alpha=5\times10^{-2}$\ (Rates shown on the
scale of $10^{-2}$)}\\
minP & 4.89 & 4.85 & 4.44 & & 4.92 & 4.84 & 4.16 & & 7.63 & 7.46 & 7.25
\\
Global & 2.73 & 2.71 & 2.72 & & 3.01 & 2.99 & 2.96 & & 2.75 & 2.74 & 2.73
\\
{KM} &5.07 & 5.11 & 4.73 & & 5.10& 5.17 & 4.81 & & 5.12
& 4.97 & 4.78
\\
SimReg-PH & 5.11 & 5.10 & 4.79 & & 5.15 & 5.11 & 4.83 & & 5.21 & 5.04 &
5.01 \\
\multicolumn{12}{c}{$\alpha=5\times10^{-3}$\ (Rates shown on the
scale of $%
10^{-3}$)} \\
minP & 6.52 & 6.22 & 5.50 & & 6.13 & 5.44 & 4.39 & & 17.09 & 17.62 & 19.21
\\
Global & 2.82 & 2.65 & 2.67 & & 2.98 & 2.74 & 2.86 & & 2.61 & 2.50 & 2.59
\\
{KM} & 5.18 & 4.86 & 4.02 & & 5.26 & 4.86 & 4.28 & & 5.40
& 5.54 & 4.96\\
SimReg-PH & 5.18 & 4.91 & 4.15 & & 5.22 & 5.03 & 4.38 & & 5.84 & 5.20 &
5.96 \\
\multicolumn{12}{c}{$\alpha=5\times10^{-4}$\ (Rates shown on the
scale of $%
10^{-4}$)} \\
minP & 8.6 & 8.1 & 6.8 & & 8.0 & 7.0 & 6.2 & & 48.4 & 47.9 & 60.3 \\
Global & 2.9 & 3.9 & 2.7 & & 4.0 & 3.2 & 2.4 & & 3.1 & 2.9 & 2.8 \\
{KM} & 5.8 & 5.2 & 2.8 & & 6.8 & 5.8 & 4.4 & & 8.4 &
4.4 & 6.6\\
SimReg-PH & 7.1 & 5.7 & 2.7 & & 6.9 & 5.4 & 3.2 & & 8.0 & 4.6 & 8.1 \\
\hline
\end{tabular*}
\tabnotetext[]{}{The type I error rates are shown on the scale of
$10^{2}$, $10^{3}$,
and $%
10^{4}$ for nominal level at 0.05, 0.005, and 0.0005, respectively. The
survival times were generated from the PH model and were analyzed using
different approaches under the PH model. The results were based on $10^5$
replications except that the results for KM were based
on $5\times10^4$ replications.}
\end{sidewaystable}

We generated the censoring time, $C_{i}$, from Unif$ ( 0,c )
$, where $c$ is uniquely chosen for each of 3
censoring rates: 15\%, 40\% and 90\%. Specifically, we set $c = 6.7$,
2.0 and 0.2 for censoring rates of 15\%, 40\% and 90\%, respectively.
The sample sizes, $n$, were 500 for the 15\% and 40\% censoring rates
under the additive and dominant modes. For the 90\% censoring rate
under the additive and dominant modes and all censoring rates under the
recessive mode, $n=1000$.
Each scenario was analyzed using SimReg (PH and/or PO), minP, Global
{and KM. SimReg-PH and KM have identical test statistics
but used different resampling approaches to obtain $p$-values. Because
both resampling approaches are asymptotically equivalent, we expect
minor differences in finite sample performance between SimReg-PH and
KM.} In all analyses, the causal loci were excluded.

\subsection{Results of type I error analyses}

We first examined the performance of the proposed SimReg-PH model.
Table~\ref{tab3} displays the type I error rates of different methods when the
survival times were generated from the PH model. The results were based
on $10^5$ replications {except that KM was based on
$5\times10^4$ replications due to computational cost.}
In each replication, the $p$-values of SimReg-PH were obtained from
$5\times10^5$ resamplings.
We report the type I error rates evaluated at the nominal levels of
$5\times10^{-2}$, $5\times10^{-3}$ and $5\times10^{-4}$.
The type I error rates obtained by SimReg-PH remained around the
nominal levels. However, the deviations were larger for $\alpha=5\times
10^{-4}$, mainly due to fewer resampled statistics observed on the
extreme tail.
In particular, for the low censoring proportion (i.e., 15\%), the type
I error rates were slightly inflated, while for the high censoring
proportion (i.e., 90\%), the type I error rates became a little
conservative when $\alpha=5\times10^{-4}$.
Nevertheless, the overall results suggest that the SimReg-PH test
maintained an appropriate size, which confirms the validity of the
derived null distribution of the test statistic, $Q_{n}$.
{As expected, the type I error rates obtained by KM were
very similar to SimReg-PH.}
The type I error rates obtained by the Global test are overly
conservative; similar behavior has been reported in the
literature [\citet{ZhoChe11}]. The minP method had correct type
I error rates under additive and dominant modes. However, the method
had inflated
type I error rates under the recessive mode, and the inflation was more
severe with smaller $\alpha$.
Under the recessive mode, the Bonferroni corrected $p$-values obtained by
replacing $K_{\mathrm{eff}}$ with the total number of SNPs
also yielded inflated type I error rates. Specifically, the empirical
type I error rates for 15\%, 40\% and 90\% censoring are
(0.0627, 0.0687, 0.0608) for $\alpha=5 \times10^{-2}$,
(0.0145, 0.0152, 0.0163) for $\alpha=5 \times10^{-3}$ and
(0.0042, 0.0041, 0.0053) for $\alpha=5 \times10^{-4}$, respectively.
The anti-conservation appears to be somewhat related to the rare
recessive loci; when the rare loci are excluded from the analysis,
the empirical type I error rate became closer to the nominal level
(data not shown). However, such an exclusion strategy might give
uninformative results because the relevant signals were excluded from
the analysis.

Table~\ref{tab4} displays the type I error rates when the survival times were
generated from the PO model.
Both SimReg-PH and SimReg-PO were implemented. The results were based
on $5000$ replications, and the type I error rates were calculated at
the nominal level of 0.05. The type I error rates for both SimReg-PO
and SimReg-PH were close to 0.05 independent of the inheritance mode
and the censoring proportions. These results show the validity of the
SimReg-PO tests and the robustness of the SimReg-PH tests.
{Though not performed, KM is expected to have the same
robustness as SimReg-PH.}
As seen previously, the Global test had conservative type I error
rates, but the magnitude of conservation is less than that seen in
Table~\ref{tab3}. The minP method yielded slightly inflated type I error rates
for the additive and dominant modes with low censoring proportion
(i.e., 15\%). As before, it yielded inflated type I error rates for the
recessive mode.

%
\begin{sidewaystable}
\tabcolsep=0pt
\tablewidth=\textwidth
\caption{Type I error rates for survival time generated from
the PO model}\label{tab4}
\begin{tabular*}{\textwidth}{@{\extracolsep{\fill}}lccccccccccc@{}}
\hline
& \multicolumn{3}{c}{\textbf{Additive}} & &
\multicolumn{3}{c}{\textbf{Dominant}} &&
\multicolumn{3}{c}{\textbf{Recessive}}\\[-6pt]
& \multicolumn{3}{c}{\hrulefill} &&
\multicolumn{3}{c}{\hrulefill} &&
\multicolumn{3}{c@{}}{\hrulefill}\\
 & \multicolumn{1}{c}{\textbf{15\%}} & \multicolumn{1}{c}{\textbf{40\%}} &
\multicolumn{1}{c}{\textbf{90\%}} & & \multicolumn{1}{c}{\textbf{15\%}} &
\multicolumn{1}{c}{\textbf{40\%}} & \multicolumn{1}{c}{\textbf{90\%}} & &
\multicolumn{1}{c}{\textbf{15\%}} & \multicolumn{1}{c}{\textbf{40\%}} &
\multicolumn{1}{c@{}}{\textbf{90\%}}\\
 & \multicolumn{1}{c}{$\bolds{(n=500)}$} &
\multicolumn{1}{c}{$\bolds{(n=500)}$} &
\multicolumn{1}{c}{$\bolds{(n=1000)}$} & &
\multicolumn{1}{c}{$\bolds{(n=500)}$} &
\multicolumn{1}{c}{$\bolds{(n=500)}$} &
\multicolumn{1}{c}{$\bolds{(n=1000)}$} & &
\multicolumn{1}{c}{$\bolds{(n=1000)}$} &
\multicolumn{1}{c}{$\bolds{(n=1000)}$} &
\multicolumn{1}{c@{}}{$\bolds{(n=1000)}$} \\
\hline
\multicolumn{1}{@{}l}{minP} & 0.0584 & 0.0548 & 0.0444 & & 0.0560 & 0.0492 &
0.0410 & & 0.0798 & 0.0774 & 0.0740 \\
\multicolumn{1}{@{}l}{Global} & 0.0358 & 0.0320 & 0.0256 & & 0.0352 &
0.0332 &
0.0266 & & 0.0342 & 0.0312 & 0.0250 \\
\multicolumn{1}{@{}l}{SimReg-PH} & 0.0488 & 0.0496 & 0.0464 & & 0.0492 & 0.0504
& 0.0478 & & 0.0526 & 0.0518 & 0.0480 \\
\multicolumn{1}{@{}l}{SimReg-PO} & 0.0446 & 0.0486 & 0.0468 & & 0.0452 & 0.0496
& 0.0508 & & 0.0544 & 0.0492 & 0.0490 \\
\hline
\end{tabular*}
\tabnotetext[]{}{The survival times were generated from the
PO model and were analyzed using
different approaches under the PO model or the PH model (to examine the
impact of model misspecification). The results were based on 5000
replications.}
\end{sidewaystable}
%
\begin{figure}

\includegraphics{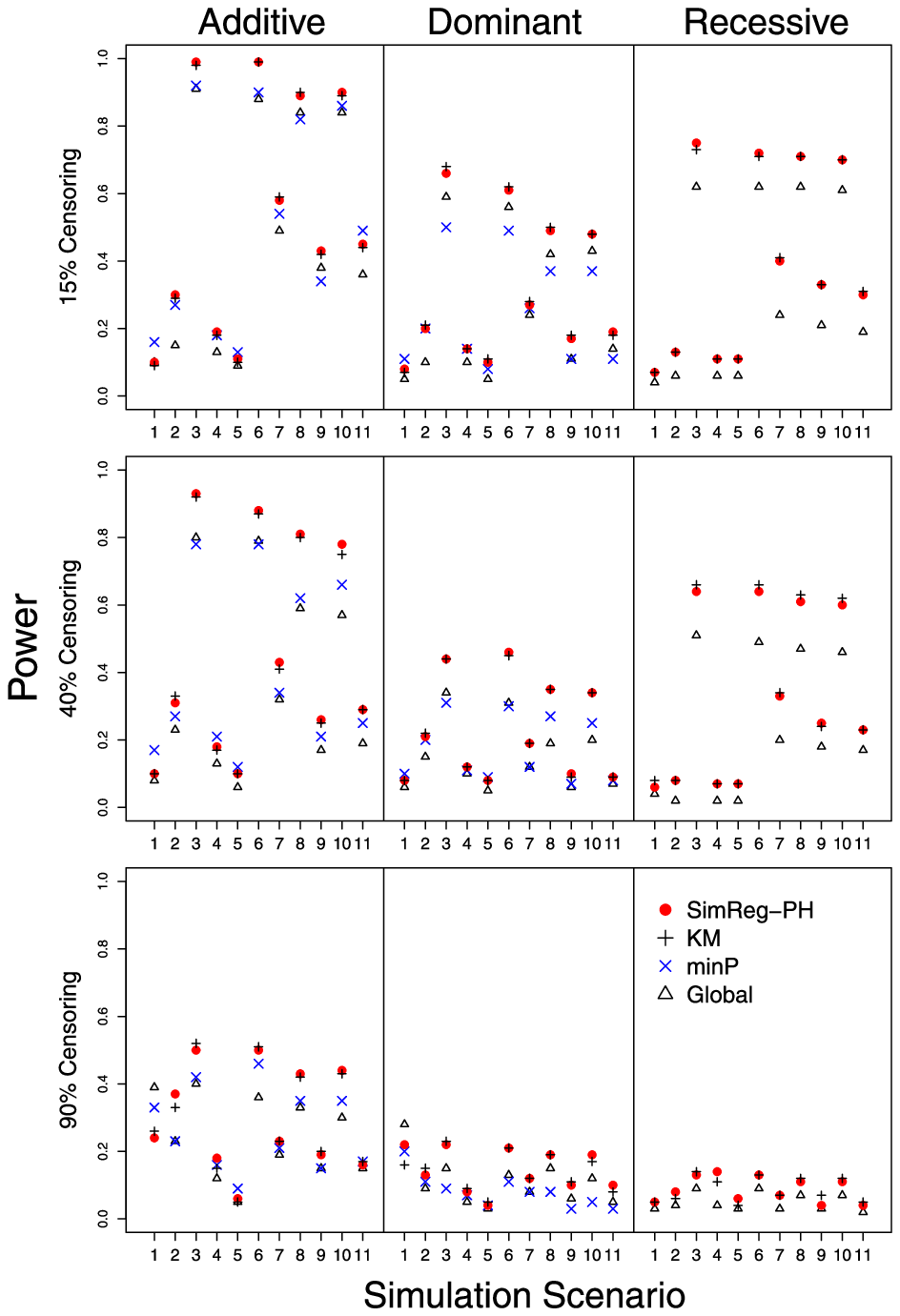}

\caption{Power when the survival times are generated from the PH model.}
\label{fig.PHpower}
\end{figure}

\subsection{Results of power analyses}

{The power analyses were performed using the settings specified in
Table~\ref{tab2}. The results were based on 100 replications under each scenario.
Figure~\ref{fig.PHpower} shows the power when the survival times
were generated from the PH model.
We first consider the additive mode. When one large-effect causal locus
has low MAF and low LD with the other markers (e.g., Scenarios 1, 5 and~11), minP tends to have the highest power independent of the censoring
proportion. The good performance of minP is not unexpected in these
scenarios because the overall association of the gene was driven by a
single large-effect locus, for which the majority of the other SNPs did
not carry much information. As a result, there is no power gain when
borrowing information from other SNPs, which is what SimReg-PH does.
However, the power gain of minP over other methods generally diminishes
as the number of causal loci increases (e.g., Scenarios~5 to~10). In
scenarios where the marker set is not dominated by a single causal
locus of low MAF and low LD, SimReg-PH showed comparable or higher power.
{As expected, KM had near identical power as SimReg-PH in
all scenarios.}
In most scenarios, the Global test produces the least amount of power
largely due to the over-conservative test size. 
The overall performance under the dominant mode has a similar trend to
that of the additive mode. However, the power of SimReg-PH is
comparable to or better power than the power of minP in more cases
under the dominant mode than under the additive mode. For the recessive
mode, SimReg-PH appears to have better power than the Global test. We
did not perform a power analysis for minP because the type I error
rates were inflated.

\begin{figure}

\includegraphics{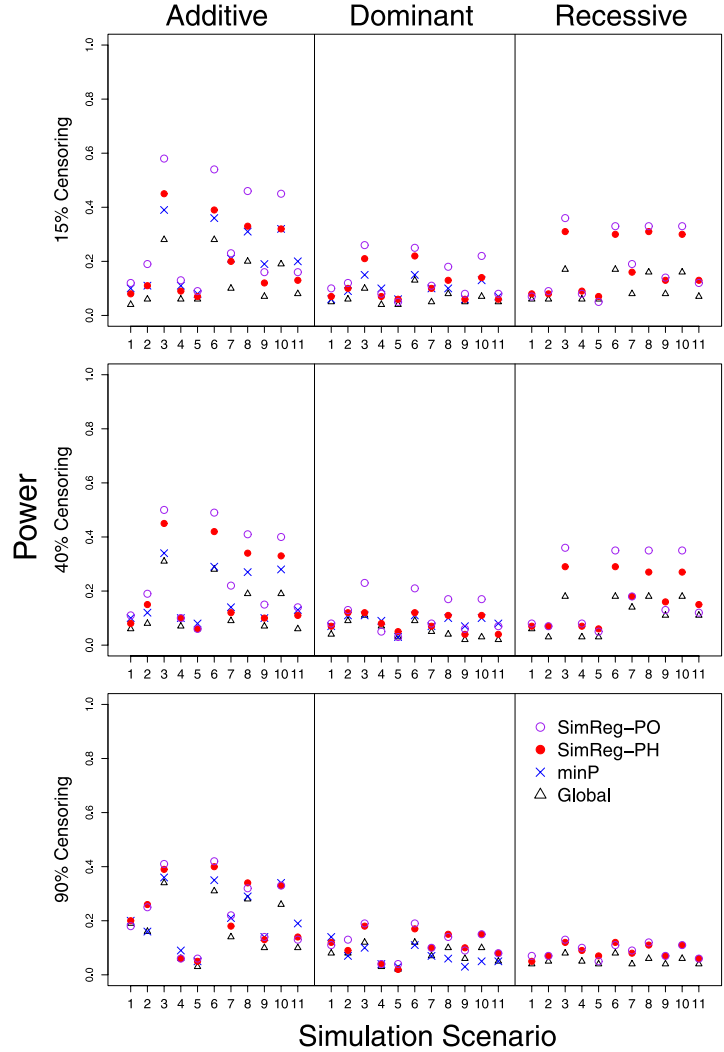}

\caption{Power when the survival times are generated from the PO model.}
\label{fig.POpower}
\end{figure}

Figure~\ref{fig.POpower} shows the power performance when the survival
times were generated from the PO model. The power obtained using
SimReg-PO is similar to or better than the power obtained using
SimReg-PH, indicating that efficiency is gained when the correct model
is used. The power gain of SimReg-PO is more substantial when the
censoring proportion is low to medium, and the power is comparable to
or better power than the minP power in some of those difficult cases,
such as Scenario 1.}

\section{Discussion}

In this work we extended the similarity regression (SimReg)
approaches, which have shown effectiveness in modeling marker-set
effects on binary and continuous outcomes, to survival models to
facilitate the assessment of gene or pathway effects on drug
responses. The genetic effect is evaluated by assessing the
association between the IBS status of a pair of individuals and the
covariance of their survival times.
We derived the equivalence between the similarity survival
regression and a random effects model. The equivalence facilitates the
derivation of the score test statistics and unifies the
current variance component-based methods. {Specifically, the KM
approach [\citet{Linetal11}] is equivalent to
SimReg-PH when the same kernel function} is used to quantify the
genetic similarity, $S_{ij}$. The Global test [\citet{Goeetal05}]
can be viewed as a special case of SimReg-PH with $S_{ij}=\sum_m \mbox
{\textsc{g}}_{m,i}^T \mbox{\textsc{g}}_{m,j}$ (i.e., the linear
kernel). However, the results of Global and SimReg-PH with linear
kernels may be different because different approaches were used to
derive the asymptotic distributions of the test statistics.
Compared to these existing gene-based approaches, our proposed method
has the
generality to incorporate a variety of risk models in the class of
linear transformation models{, and we explicitly
constructed the SimReg tests under the PH model and the PO model in
this work. We also proposed a resampling approach to obtain the
$p$-values that improves the computational efficiency. Finally,} we
showed that the derived inference procedure is
robust against the misspecification of the risk model, which is an
attractive feature because the underlying risk model is often unknown.
Through simulations, we
showed that the power of the SimReg method is comparable to or higher
than the power of the minP and
Global methods across various scenarios. We also verified that the
SimReg-PH test statistics remain valid even if the risk model is
misspecified.

In the data application on the VISP study, we illustrated how SimReg can
be used to search for genes or pathways that are associated with
a time-to-event outcome and confirmed previous findings using this
gene-based approach with statistical significance. Although we
focused our method development and demonstrated its utility based on
pharmacogenetics studies, the proposed method is applicable to other
genetic clinical researches or observational studies with
time-to-event outcomes.
{For a pharmacogenetic study with sample sizes 1000, such
as the VISP trial, 1000 runs of the SimReg analyses took $\le$1 hour to
complete on an Intel Xeon 3.33 GHz machine with 12 Gb RAM using one
processing core. We expect a gene-based whole genome analysis on $\sim
$20K genes should be completed in a day using a comparable computing
facility.}
We implemented the
proposed methods in R and made it available at the authors' websites.
We are incorporating the software into the \texttt{SimReg} R package.

Motivated by the data application where the risk variant acted recessively,
we further investigated the behavior of the gene-based approaches under
different modes of inheritance. We found that all of the studied gene-based
methods performed appropriately under the additive and dominant modes. However,
caution should be used when performing the minimal $p$-value approaches under
the recessive mode; the minimal $p$-value approaches had severely
inflated type I error rates.
The inflation might be related to the extremely
rare recessive loci. However, excluding those rare recessive loci is
suboptimal because the important signals can be artificially removed
and lead
to power loss. In contrast, the global test and the similarity regression
are not vulnerable to such a situation and appear to be more suitable options
given their reasonable performance under the recessive mode.

This work focused on assessing the genetic main effect on drug
response in an effort to understand how individual variation affects
drug efficacy and toxicity. In pharmacogenetics and personalized
medicine, one major topic is to study if the genetic effects are
modified by treatments and how the effects differ across treatment
options. As observed in the VISP genetic studies, the effect of \emph
{TCN2} on
recurrent stroke risk is restricted to the low-dose treatment.
An analysis stratified by treatments allows for the evaluation of such
heterogeneous effects between different treatment groups, but its
efficiency can be further improved by incorporating the gene-treatment
interaction in the regression model. Such an extension is not
straightforward because the calculation of the score test requires the
variance component estimates for the genetic main effect under a
mixed-effects survival model. We are developing further extensions of
SimReg to incorporate interaction effects.

\begin{appendix}\label{app}
\section*{Appendix}

\textit{Derivation of the score statistic $Q_{n}$}:
Given the working mixed model $H(T_{i})=\gamma^{T}X_{i}+g_{i}+\varepsilon
_{i}^{\ast}$, the log-likelihood function of the observed data can be
written as
\begin{eqnarray*}
&&l_{n}(\gamma,H,\tau)
\\
&&\qquad=\log\int\cdots\int\prod_{i=1}^{n}\bigl[
\lambda\bigl\{H(\tilde{T%
}_{i})-\gamma^{T}X_{i}-g_{i}
\bigr\}\bigr]^{\delta_{i}}\\
&&\hspace*{96pt}{}\times e^{-\Lambda\{H(\tilde{T}%
_{i})-\gamma^{T}X_{i}-g_{i}\}}f_{G}(g_{1},
\ldots,g_{n})\,dg_{1}\cdots dg_{n},
\end{eqnarray*}
where {$\lambda$ and $\Lambda$ are the specified hazard
and cumulative hazard functions of $\varepsilon
_{i}^{\ast}$}, and $f_{G}(g_{1},\ldots,g_{n})$ is the joint density of
$g_{1},\ldots,g_{n}$,
that is, a multivariate normal density with mean 0 and variance--covariance
matrix $\tau S$. {Consider the variable transformation
$(g^*_{1},\ldots,g^*_{n})' = \tau^{-1/2}S^{-1/2}(g_{1},\ldots,g_{n})'$, where
$S = S^{1/2}S^{1/2}$. Then $(g^*_{1},\ldots,g^*_{n})'$ follows a standard
multivariate normal distribution. The result leads to
\begin{eqnarray*}
&&l_{n}(\gamma,H,\tau)
\\
&&\qquad=\log\int\cdots\int\prod_{i=1}^{n}\bigl[
\lambda\bigl\{H(\tilde{T%
}_{i})-\gamma^{T}X_{i}-
\tau^{1/2}S^{1/2}g^*_{i}\bigr\}\bigr]^{\delta
_{i}}\\
&&\hspace*{96pt}{}\times e^{-\Lambda\{H(\tilde{T}%
_{i})-\gamma^{T}X_{i}-\tau^{1/2}S^{1/2}g^*_{i}\}}
\\
& &\hspace*{64pt}\qquad\quad{} \times f^*_G\bigl(g^*_{1},\ldots,g^*_{n}
\bigr)\,dg^*_{1}\cdots dg^*_{n},
\end{eqnarray*}
where $f^*_G$ is the density for the standard multivariate normal
distribution.} After some algebra, we have
\begin{eqnarray*}
&&\frac{1}{n}\frac{\partial l_{n}(\gamma,H,\tau)}{\partial\tau} \bigg\vert%
_{\tau=0}
\\
&&\qquad=\frac{1}{2n}(r_{1},\ldots,r_{n})S(r_{1},
\ldots,r_{n})^{T}\\
&&\qquad\quad{}+\frac{1}{2n}%
\sum
_{i=1}^{n} \biggl[ \frac{\ddot{\lambda}(e_{i})\lambda(e_{i})-\{\dot{%
\lambda}(e_{i})\}^{2}}{\lambda^{2}(e_{i})}-\dot{
\lambda}(e_{i}) \biggr] s_{i}^{T}s_{i},
\end{eqnarray*}
where $e_{i}=H(\tilde{T}_{i})-\gamma^{T}X_{i}$, $\ddot{\lambda}(\cdot
)$ is
the second derivative of $\lambda(\cdot)$, $s_{i}$ is the $i$th row
of the
matrix $S^{1/2}$ and $r_{i}=\int_{0}^{\infty}\dot{\lambda}\{H(t)-\gamma
^{T}X_{i}\}/\lambda\{H(t)-\gamma^{T}X_{i}\}\,dM_{i}(t;\break  \gamma,H)$.
{The equality in the above equation is obtained by first
taking the derivative of $ l_{n}(\gamma,H,\tau)$
with respect to $\tau$ and then deriving its limit as $\tau\rightarrow
0$ using L'H\^{o}pital's rule.}
Note that the first term on the right-hand side
of the above equation is nonnegative, and the second term converges in
probability to a constant as $n$ goes to infinity. In addition, under the
null hypothesis, $r_{i}$'s have expectation of 0 at the true values of $
\gamma$ and $H$ because they are martingale integrations. Therefore,
if the
null hypothesis is correct, the first term in the summation should be
close to 0.
This {result} motivates us to consider a score test and
reject the null hypothesis
when the score $(1/n)\partial{l_{n}(\hat{\gamma},\hat{H},\tau)/
\partial\tau}\vert_{\tau=0}$ is bigger than some value, where $\hat{%
\gamma}$ and $\hat{H}$ are the estimates of $\gamma$ and $H$, respectively,
under the null model. It is asymptotically equivalent to consider the
test statistic $Q_{n}=n^{-1}(\hat{r}_{1},\ldots,\hat{r}_{n})S(\hat
{r}_{1},\ldots,%
\hat{r}_{n})^{T}$ and reject the null hypothesis when $Q_{n}>c_{\alpha}$,
where $c_{\alpha}$ is the critical value for a level-$\alpha$ test.

\textit{Null distribution of the score statistic $Q_{n}$}:
Here, we consider the estimators $\hat{\gamma}$ and $\hat{H}%
(\cdot)$ obtained via the martingale-based estimating equations for
the standard linear transformation model [\citet{CheJinYin02}]
under the null hypothesis. Note that $Q_{n}=(n^{-1/2}\sum_{i=1}^{n}\hat
{r}%
_{i}s_{i})^{T}(n^{-1/2}\sum_{i=1}^{n}\hat{r}_{i}s_{i})$. Let $\gamma_0$
and $H_0$ denote the true values of $\gamma$ and $H$, respectively, in
the null model. Based on
the derivations given in \citet{CheJinYin02}, we have the
following asymptotic representations:
\begin{eqnarray*}
\sqrt{n}(\hat{\gamma} - \gamma_0) &=&-A^{-1}
\frac{1}{\sqrt{n}}\sum_{i=1}^n\int
_0^\infty\bigl\{X_i -
\mu_{X}(t)\bigr\}\,dM_i(t;\gamma_0,H_0)
+ o_p(1),
\\
\sqrt{n}\bigl\{\hat{H}(t) - H_0(t)\bigr\} &=& -b_1(t)^TA^{-1}
\frac{1}{\sqrt{n}}\sum_{i=1}^n\int
_0^\infty\bigl\{X_i -
\mu_{X}(t)\bigr\}\,dM_i(t;\gamma_0,H_0)
\\
& &{} + \frac{1}{\sqrt{n}}\sum_{i=1}^n\int
_0^t \phi(t,s) \,dM_i(s;\gamma
_0,H_0) + o_p(1).
\end{eqnarray*}
The definitions of $A$, $\mu_X(\cdot)$, $b_1(\cdot)$ and $\phi(\cdot,\cdot)$ can be found in \citet{CheJinYin02}. By the Taylor
expansion, we can show that
\[
\frac{1}{\sqrt{n}}\sum_{i=1}^{n}
\hat{r}_{i}s_{i}\rightarrow_{d}N(0,\Sigma),
\]
as $n\rightarrow\infty$, where $\Sigma=E(\psi_{i}\psi_{i}^{T})$ and
\begin{eqnarray*}
\psi_{i}&=& \biggl[\delta_i\frac{\dot{\lambda} \{H_0(\tilde{T}_i)-\gamma
_0^TX_{i}\}}{\lambda\{H_0(\tilde{T}_i)-\gamma_0^TX_{i}\}} - \lambda
\bigl\{ H_0(\tilde{T}_i)-\gamma_0^TX_{i}
\bigr\} \biggr]s_{i}
\\
& &{}-(B_{1}-B_{2})A^{-1}\int_{0}^{\infty}
\bigl\{X_{i}-\mu_X (t)\bigr\}\,dM_{i}(t;
\gamma_0,H_0)
\\
& &{}-\int_{0}^{\infty}b_2(t)\,dM_{i}(t;
\gamma_0,H_0).
\end{eqnarray*}
Here
\begin{eqnarray*}
B_1 &=& \lim_{n \rightarrow\infty}\frac{1}{n}\sum
_{i=1}^n s_iX_i^T
\bigl[\dot{\lambda}\bigl\{H_0(\tilde{T}_i)-
\gamma_0^TX_{i}\bigr\}
\\
& &\hspace*{73pt}{}-\delta_i\bigl(\ddot{\lambda}\bigl\{H_0(\tilde{T}_i)-\gamma
_0^TX_{i}\bigr\}\lambda\bigl\{H_0(\tilde{T}_i)-\gamma_0^TX_{i}\bigr\} \\
&&\hspace*{159pt}{}- \bigl(\dot{\lambda
}\bigl\{H_0(\tilde{T}_i)-\gamma_0^TX_{i}\bigr\}\bigr)^2\bigr)\\
&&\hspace*{172pt}/\bigl(\lambda\bigl\{H_0(\tilde
{T}_i)-\gamma_0^TX_{i}\bigr\}\bigr)^2 \bigr],
\\
B_2 &=& \lim_{n \rightarrow\infty}\frac{1}{n}\sum
_{i=1}^n s_i b_1(
\tilde{T}_i)^T \bigl[\dot{\lambda}\bigl
\{H_0(\tilde{T}_i)-\gamma_0^TX_{i}
\bigr\}
\\
& &\hspace*{93pt}{}-\delta_i\bigl(\ddot{\lambda}\bigl\{H_0(\tilde{T}_i)-\gamma
_0^TX_{i}\bigr\}\lambda\bigl\{H_0(\tilde{T}_i)-\gamma_0^TX_{i}\bigr\} \\
&&\hspace*{177pt}{}-
\bigl(\dot{\lambda
}\bigl\{H_0(\tilde{T}_i)-\gamma_0^TX_{i}\bigr\}\bigr)^2\bigr)\\
&&\hspace*{189pt}{}/
{\bigl(\lambda\bigl\{H_0(\tilde
{T}_i)-\gamma_0^TX_{i}\bigr\}\bigr)^2} \bigr],
\\
b_2(t) &=& \lim_{n \rightarrow\infty}\frac{1}{n}\sum
_{i=1}^n s_i I(
\tilde{T}_i \ge t)\\
&&\hspace*{47pt}{}\times\phi(\tilde{T}_i,t) \bigl[\dot{
\lambda}\bigl\{H_0(\tilde{T}_i)-\gamma_0^TX_{i}
\bigr\}
\\
& &\hspace*{98pt}{}-\delta_i\bigl(\ddot{\lambda}\bigl\{H_0(\tilde{T}_i)-\gamma
_0^TX_{i}\bigr\}\lambda\bigl\{H_0(\tilde{T}_i)-\gamma_0^TX_{i}\bigr\}
\\
&&\hspace*{182pt}{}- \bigl(\dot{\lambda
}\bigl\{H_0(\tilde{T}_i)-\gamma_0^TX_{i}\bigr\}\bigr)^2\bigr)\\
&&\hspace*{194pt}/{\bigl(\lambda\bigl\{H_0(\tilde
{T}_i)-\gamma_0^TX_{i}\bigr\}\bigr)^2} \bigr].
\end{eqnarray*}

Therefore, $Q_{n}$ converges in distribution to a weighted $\chi^{2}$
distribution: $\sum_{k=1}^{d}\xi_{k}\chi_{1,k}^{2}$, where $\chi
_{1,1}^{2},\ldots,\chi_{1,d}^{2}$ are $d$ independently and identically
distributed $\chi^{2}$ random variables with degree freedom of 1, and $
\xi_{1},\ldots,\xi_{d}$ are the $d$ nonzero eigenvalues of the matrix
$\Sigma$. To obtain the critical value, $c_{\alpha}$, of the limiting
weighted $\chi^{2}$ distribution, we use a numerical method. Specifically,
we first obtain a consistent estimator, $\hat{\Sigma}$, of $\Sigma$
using the
usual plug-in method and compute the nonzero eigenvalues $\hat{\xi
}_{1},\ldots,\hat{\xi}_{d}$ of the matrix $\hat{\Sigma}$. Next we
generate a
large set (e.g., 10,000 sets) of independent and identically distributed
random variables $\chi
_{1,1}^{2},\ldots,\chi_{1,d}^{2}$. For each set of $\chi^{2}$ random
variables, we compute $\sum_{k=1}^{d}\hat{\xi}_{k}\chi_{1,k}^{2}$. We can
then estimate $c_{\alpha}$ by the upper $\alpha$-quantile of $%
\sum_{k=1}^{d}\hat{\xi}_{k}\chi_{1,k}^{2}$'s and the associated
$p$-value of
the score test can be computed accordingly.
\end{appendix}
\section*{Acknowledgments}
The authors thank Dr. Mich\`{e}le M. Sale and Dr. Bradford B.
Worrall for providing the data. They thank Dr. Chen-Hsin Chen
for his insightful discussion and helpful suggestions to improve the
work. They also thank Dr. Shannon Holloway for her constructive input
{and proofreading} to improve the manuscript and
assistance in creating the software package.

\printaddresses

\end{document}